# First-Principles Calculations of Glycine Formation on Cu(110) Surface

*Po-Tuan Chen*

Center for Condensed Matter Sciences, National Taiwan University, Taipei 106, Taiwan

E-mail address: r92222019@ntu.edu.tw

**ABSTRACT**

The geometrical structures of singlet or triplet $NCH_3$ molecules adsorbing on Cu(110) surface with presence of Cu adatoms have been determined by first-principle calculation method. The two distinguishable structures match the patterns of experimental scanning tunneling microscopy (STM) image, so-called zigzag and rectangle structures. The singlet $NCH_3$ molecules arrange as zigzag, when the triplet $NCH_3$ molecules array as rectangle. Since $NCH_3$ can be found in interstellar environment and its radicals are chemically active, we arise a question whether this system can be utilized to study surface chemical reaction of interstellar medium (ISM). Therefore, the potential surface of $NCH_3$ isomerization to $HNCH_2$ subsequently binding with CO on Cu(110) surface has been calculated using density functional theory (DFT). The calculation results that $HNCH_2$-CO can be produced on triplet state. However, the reaction is endothermic which is assumed to be occurred at warmer regions of solar systems or in specific planets having terrestrial heat (like Earth).

# INTRODUCTION

Including well-characterized surface processes into modeling chemical reactions of interstellar medium (ISM) molecules is one of challenges to extensively study astrochemistry [1]. The ISM consists of gas and dust between the stars. Much of ISM material has been synthesized from precursor atoms which has been ejected by old and dying stars. It is retained in different environments showing large ranges in temperature ($10$–$10^4$ K) and densities ($10^0$–$10^8$ number of particles cm$^{-3}$). Very hot region at very low density contains with H$^+$ ion or with H radical, called HII region; as well as, interstellar clouds are characterized by very low temperatures (10–100 K) and high densities ($10^2$–$10^8$ number of particles cm$^{-3}$). The presence of species in the space surrounding stars is investigated using spectroscopy methods. ISM is filled mainly with H$_2$ gas, about 10% He atoms, and < 0.1% of atoms such as C, N, and O. Cold gas phase chemistry can efficiently form simple species such as CO, N$_2$, O$_2$, C$_2$H$_2$ and C$_2$H$_4$, HCN, and simple carbon chains [2]. Copper, Gallium, Germanium, and Krypton have been detected in the ISM [3] which are often incorporated in dusts. These heavy elements have been synthesized in massive stars [4]. Roughly 1% of the mass is in micron-sized dust grains. In fact, universe acts as a giant factory synthesizing molecules. Interstellar molecules can be synthesized via gas-phase processes and via reactions on the surface of the dust particles. Additionally, the chemistry in ISM and the production of molecules can be enriched by ultraviolet (UV) irradiation and cosmic rays. Cosmic rays can penetrate throughout the cloud and ionize H$_2$ molecules; the energetic electrons from these molecules can in turn excite H$_2$ to higher electronic states. These excited H$_2$ molecules subsequently decay by emitting fluorescence [5,6]. These secondary photons can photoexcite and photoionize neutral molecules. Combined information from astronomical observations of infrared and radio from interstellar species, laboratory spectroscopy, and supporting theoretical models allows us to understand the formation and distribution of species in the dense ISM. [7]

Well over one hundred different molecules have been identified in interstellar and circumstellar regions, most of them are organic species [8]. We can point out the ubiquitous presence of cyanides. At the other end of Universe, the pair of isomers HNC / HCN has also been detected. This fact is fascinating for those who are interested in the life-origin problem [9]. The existence of molecules of biological interest in ISM has attracted great attention. Direct spectroscopic observations evidence the presence of prebiotic species such as aminoacetonitrile. The glycine-like molecule aminoacetonitrile was discovered by the Max Planck Institute for Radio Astronomy [10]. Large quantities of the organic material necessary to initiate prebiotic chemistry could have been delivered to the early Earth by the impacts of comets and asteroids [11,12,13]. Simple interstellar organics may have also acted as the precursors of the more complex organics found in meteorites, formed through aqueous processing on their parent bodies [8]. We could in principle trace the origin of Earth's prebiotic chemistry back to the parent interstellar cloud. If more complex organics exist in interstellar space [14], then they may have "jump started" prebiotic chemistry on the early Earth by meteors. Glycine is the simplest amino acid, and due to the significant astrobiological implications that suppose its detection, the search for it in the ISM, meteorites, and comets is intensively investigated. About ten years ago, the detection of glycine in the ISM has been debated [15,16]. In 2009, glycine sampled from comet by the NASA spacecraft Stardust was confirmed. This is a definite discovery of glycine outside the Earth, however, glycine was identified in the Murchison meteorite in 1970 [17].

Several experiments carried out in terrestrial laboratories based on UV irradiation at cold temperature and ultra-vacuum simulated interstellar conditions, and resulted in the successful synthesis of amino acids and other organic molecules [18]. In particular, glycine can be formed in a laboratory under conditions simulating the ISM ones. For instance, UV irradiation or electron bombardment of $NH_3$ and $CH_3COOH$ or $CH_3NH_2$ and $CO_2$ ices brings about the glycine molecule [19,20,21]. Quantum mechanical methods have also been adopted to study a variety of possible reaction mechanisms by exploring the potential energy surfaces. The reaction of $HNCH_2$, CO and $H_2O$ for the formation of glycine have been investigated using quantum chemical methods based on density functional theory (DFT) [22]. This reaction proceeds through a concerted mechanism with reasonably large barriers for the cases with one and two water molecules as reactants. With an increasing number of water molecules as the reactants, when the numbers of water molecules are three or more than three, the barrier height reduced so drastically that the transition states were more stable than the reactants. On the other hand, the formation of glycine in the presence of a radical water ice as well as of an ionized water ice has been simulated by DFT [23]. The excited water ices are mimicked the effect of UV and cosmic rays irradiation. The COOH radicals couple with incoming $HNCH_2$ molecules to form the $NHCH_2COOH$ radical. Nonetheless, when $H_3O^+$ exists one proton may be barrierlessly transferred to $HNCH_2$ to produce $NH_2CH_2^+$. The latter may react with the COOH radical to give the $NH_2CH_2COOH^+$ glycine radical cation. The results provide quantum chemical evidence that defects formed on water ices due to the harsh-physical conditions of the ISM may trigger reactions of astrochemical interest.

To our best knowledge, a question as to whether dust grains consisting of heavy elements can play a role in catalysis has still been unanswered. Current uncertainties regarding chemistry of dust are, therefore, so great. A prior question may arise: if there exists a well-characterized surface adsorption system proper for studying astrochemistry. In the light of this purpose, we introduce a system of $NCH_3$ adsorbed on Cu(110) surface. $NCH_3$ is a Jahn–Teller active and short-lived specie which quickly transforms into $HNCH_2$. $NCH_3$ has been experimentally detected in Titan atmosphere while $HNCH_2$ has found in interstellar cloud [24]. These cyanide-based species are of considerable interests because it is connected to the origin of life [22,23]. The stable existence of the $NCH_3$ on Cu(110) in laboratory has been identified by vibrational spectra, temperature-programmed desorption, ultraviolet photoemission spectra, and scanning tunneling microscopy (STM) [25,26,27]. Our hypothesis is following: on Cu surface, singlet $NCH_3$ is most likely to be excited upon irradiation of light, intersystem crossing may undergo to produce triplet $NCH_3$ and subsequently the active $NCH_3$ can transform to $HNCH_2$, which, in turn, easily binds with CO to form a central fragment of glycine. In this paper, we used the first principle simulation method to explore the possibility of whether a promising reaction of glycine formation from the reactant $NCH_3$ can take place.

**METHOD**

According to the previous STM study [28], NCH$_3$ species aggregate into islands with two molecules in each $p(2\times3)$ surface unit cell. For models with reconstructed Cu surface, each unit cell contains additional four Cu adatoms at (110) surface hollow sites. In our calculation the $p(2\times3)$ unit cell has an eight-layer Cu slab and each layer contains six Cu atoms. The structure of the upper six Cu layers was fully relaxed. Between the Cu slabs, a vacuum gap of greater than 10 Å was used. Then, structure geometry optimization was performed by DFT calculations using Vienna ab initio simulation package (VASP) [29,30,31] with the PW91 generalized gradient approximation [32]. Calculations were spin-polarized with an energy cut-off of 400 eV and a 5×5×1 Monkhorst–Pack grid [33].

Climbing nudged elastic band (cNEB) [34,35] method was used to determine the reaction processes of glycine formation. The cNEB method is a modification to the NEB method in which the highest energy image is driven up to the saddle point; it will be at the exact saddle point. While performing cNEB, we reduced the Cu slab to two layers in an unit cell and the k-points to 1x1x1 Monkhorst–Pack grid. The energies which we will discuss bellow are electronic energy, i.e. without temperature effect.

STM experiments were conducted using a variable-temperature microscope VT-STM (Omicron) housed in a home-built ultrahigh vacuum (UHV) chamber and gas dosing facility. The UHV system had a base pressure of ~5 × 10$^{-11}$ torr. The Cu(110) crystal was cleaned through repeated sputter/anneal cycles (Ne$^+$, 1000 eV, 870 K). Methylnitrene radicals on Cu(110) were obtained through the dissociation of adsorbed azomethane molecules at room temperature [25]. STM imaging of NCH$_3$ was performed with the sample held at room temperature.

## RESULTS AND DISCUSSION

### A. Singlet and triplet NCH$_3$ on Cu(110) surface

Gas phase NCH$_3$ adopts a C$_{3v}$ symmetry and a triplet state with two unpaired electrons, forming a diradical. Yarkony et al. [36] reported theoretical results of the electron configuration of a triplet NCH$_3$, i.e., $1a_1^2 2a_1^2 3a_1^2 4a_1^2 1e^4 5a_1^2 2e^2$. The isomerizations of triplet and singlet NCH$_3$ to HNCH$_2$ have been studied by molecular electronic structure theory [37]. The relative energy is shown in **Figure 1**. The triplet NCH$_3$ isomerization to methylenimine is predicted to be endothermic by 0.79 eV, with an activation energy of 2.28 eV. It may be concluded that ground-state methylnitrene is a relatively stable species in the absence of collisions. On the other hand, singlet HNCH$_2$ lies 2.01 eV below triplet NCH$_3$ and might be accessible if the spin-orbit coupling were substantial. There appears to be little barrier (< 0.16 eV) separating singlet NCH$_3$ from ground state HNCH$_2$ [38].

NCH$_3$ can stably adsorb on Cu(110) surface. The adsorption geometry of NCH$_3$ on Cu(110) [28] is fully determined. In the previous STM study, we observed that NCH$_3$ molecules aggregate to form p(2×3) islands. Each p(2×3) unit cell consists of two NCH$_3$ molecules that order in a zig-zag arrangement, as shown in **Figure 2(a)**. The presence of adsorbate-induced surface reconstruction is evident from pit formation; it showed that an 1:2 NCH$_3$:Cu(adatom) stoichiometry must hold. The structures were then obtained. Each NCH$_3$ molecule forms bonds to the two closest Cu adatoms alternatively along a Cu added double row, with the imaged CH$_3$ groups tilted toward the neighboring vacant (missing) rows. Ionization energy calculations which were performed using the NCH$_3$-Cu$_2$ cluster model confirm that the ground state of NCH$_3$-Cu$_2$ is a singlet instead of a triplet [39], unlike a gas phase NCH$_3$. On Cu surface, bonding to two Cu adatoms removes the intrinsic Jahn–Teller instability of NCH$_3$, and the degenerate half-filled degenerate HOMOs split into an occupied HOMO and an unoccupied LUMO. The singly ionized configuration of a singlet NCH$_3$ is either a doublet or a quartet. The calculations show that at least the first four ionization configurations are doublets.

Before performing calculations of triplet NCH$_3$ on Cu surface we make up some hypothesis. Metal surface is seen an electron-full pool. If the energy difference between the singlet ground state and the triplet excited state is not much, NCH$_3$ might exchange electron with Cu surface by irradiation stimulating. The singlet NCH$_3$ can switch to triplet and coincide with a corresponding geometrical variation. Then, we set triplet electronic state to each NCH$_3$ on Cu for DFT calculation and optimize the structure. The top and lateral views are shown in **Figure 2(b)**. The adsorbing structure of the triplet NCH$_3$ are similar with the singlet one. Each NCH$_3$ molecule forms bonds to the two closest Cu adatoms alternatively along a Cu added double row, with the CH$_3$ groups slightly tilted toward the neighboring vacant rows. One unit cell with two triplet NCH$_3$ have 0.50 eV energy higher than that with two singlet NCH$_3$. This excitation can be triggered by infrared radiant.

The diagrams and detailed geometry parameters of one singlet and one triplet NCH$_3$ with two Cu adatoms on Cu(110) surface are respectively shown in **Figure 3(a)** and **3(b)**. N bonds with two Cu adatoms and its lone pair indicates to surface, approximately to sp$^3$ type. The significant difference between them is the N-C bond tilting angle to the normal of (110) surface. The tilting angle of the singlet NCH$_3$ is 31°. And, the tilting angle of the singlet NCH$_3$ is 13°. The N-C bond stands straighter to (110). Thus, the triplet NCH$_3$ adsorbates on Cu(110) surface form different STM pattern rather than the singlet ones. Singlet NCH$_3$ adsorbing on Cu(110) is well characterized. We proposed the structure of this adsorption system based on DFT calculations and STM imaging in the previous papers [28]. Alternative arrangement of CH$_3$ groups attribute to the spots of STM image. The experimental STM of singlet NCH$_3$ molecules reveals in a zig-zag fashion. Panel (a) in **Figure 4** is the theoretical calculation results of singlet NCH$_3$ molecules comparing with the experimental zig-zag STM image. Additionally, we ever observed another unsolved STM pattern [26]; NCH$_3$ molecules arrange as "rectangle". Because a triplet NCH$_3$ has magnetic moment of 2$\mu_B$. NCH$_3$ adsorbates like molecular magnets repel to each other. Therefore, NCH$_3$ adsorbates may averagely disperse in p(2x3) unit cell to form rectangle pattern. The theoretical calculation results of triplet NCH$_3$ molecules can match with the experimental rectangle STM pattern, shown in **Figure 4(b)**.

**B. NCH$_3$ reacted with CO to form HNCH$_2$CO on Cu surface**

Quantum mechanical calculations based on DFT have been used to model the glycine formation with numbers of water molecules presence as catalyst [22] and on water-ice clusters coupled with COOH radical [23]. Radicals formed on water ices due to the harsh-physical conditions of the ISM may trigger reactions of astrochemical interest. The relevance of COOH and OH radicals facilitate chemical processes and react with HNCH$_2$ to transfer into glycine. On the other hand, irradiation excitation of NCH$_3$ adsorbates on Cu surface to triplet state bring out reactive radicals. It may also result in certain chemical reactions in ISM. In order to perform calculations of chemical reactions using DFT method for such issue, we reduce Cu slab to two layers in a unit cell and remain one $NCH_3$ molecule adsorbed on two Cu adatoms. The reactant **I** is such system plus a CO molecule, shown in **Figure 5(a)**. We assume that one H atom on $CH_3$ group transfer to N to form $HNCH_2$ [the intermedium **II** in **Figure 5(a)**]. Next, a nearby physisorbing CO migrates to bind with $HNCH_2$ and leads to the product **III** of $HNCH_2CO$ which is the middle fragment of glycine.

**Figure 5(b)** shows the potential curve of the chemical reactions on singlet state. We set $NCH_3$ binding two Cu adatoms with present of CO on Cu slab to be 0 eV of relative energy. Isomerization of $NCH_3$ to $HNCH_2$ on Cu surface is exothermal of 0.21 eV. Nitrogen of $NCH_3$ which is electronically saturated bonds with two Cu and one C and possesses an electronic lone pair. The isomerization includes an N-Cu bond breaking and an N-H bond formation. Therefore, it should overcome an energy barrier of 1.88 eV. $NCH_3$ is likely to isomer to $HNCH_2$ in gas phase, the barrier is < 0.01 eV. With CO approaching to $HNCH_2$, the relative energy get higher. Eventually, we did not obtain a local minimum for $HNCH_2CO$.

The potential curve of the chemical reactions on triplet state is shown in **Figure 5(c)**. The system with reactant of $NCH_3$ is set as 0 eV of relative energy. On triplet state $NCH_3$ possessing radicals is chemically active. The reaction of $NCH_3$ to $HNCH_2$ crosses only 0.95 eV barrier and releases 0.26 eV energy. This step needs to break an N-Cu bond and form an N-H bond, the same with on singlet state. The unique difference is that radicals facilitate the reaction. Then, CO approaches to $HNCH_2$ and binds with it. The second step has 0.70 eV barrier and endotherms 0.64 eV. The triplet $HNCH_2CO$ remains radicals may facilitate further a continuous chain reaction.

**C. Mimicking surface reactions in ISM**

About 1% of interstellar matter by mass is tied up in dust particles. The surfaces of these dust particles can provide a catalyst for chemical reactions. Therefore, one of recent tasks for astrochemists is to include surface chemical processes into their models with some assurance that the chemistry is sufficiently well understood and the model results are reliable. We introduced the system of $NCH_3$ adsorbed on Cu(110) surface into this field, since cyanide-based species are considered most plausible as prebiotic molecules. The system is manipulated in UHV, assumed as interstellar vacuum. The first step of the chemistry on interstellar dust can be depicted as a chemisorption process that a $NCH_3$ molecule strikes a Cu grain and sticks to it strongly. This process is considered possibly occurred near telluric planetary where there are abundances of metals. Preceding or following the adsorption, $NCH_3$ can be stimulated to triplet excited state by light. Radical as a reactant is chemically reactive, hence neutral radical-molecule reaction is one of plausible mechanisms for ISM. There are two mechanisms for surface chemistry: (i) the Langmuir–Hinshelwood (LH) mechanism which dominates for weakly-bound adsorbates at temperatures high enough for motion to occur, and (ii) the Eley–Rideal (ER) mechanism, in which a gas-phase species strikes a stationary adsorbate. From our former experiment [27], we have known that $NCH_3$-2Cu clusters are mobile on Cu surface at 300 K. In additional, CO is the most abundant interstellar molecule after $H_2$. A CO molecule may move in from space or migrate from another physisorbed site on Cu. The surface chemistry can go via ER or via LH mechanism. The total formation energy to produce $HNCH_2CO$ is endothermic of 0.27 eV. It is supposed to occur at high temperature region of space. They are kinetically feasible at temperatures about 3000 K. Their occurrence may take place in comets exposed to warmer regions of solar systems or in specific planets having terrestrial heat (like Earth). The requested temperatures are higher than those found in dark molecular clouds and are characteristic of hot molecular cores, of comets that have passed through warmer regions of solar systems that have experienced thermal shocks.

Cosmic dust is made of dust grains and aggregates of dust grains. These particles are irregularly shaped. The ideal adsorbed Cu surface is a compromised model, because we still have limited information of the grains so far. The rate and mechanisms of non-thermal desorption of heavy species from grain surfaces are poorly understood. The result of all these uncertainties is that modellers must make many simplifications and assumptions in their treatments. As increasing spectroscopic observations of ISM dusts and progressive development of surface technology in future, astrochemists and surface scientists can confer much appropriate systems.

**CONCLUSION**

Our paramount purpose is to model astrochemistry of dust surface by a promising well-characterized surface system. Chemical combination of two ISM molecules with assistance of dust as catalyst and light stimulation in low density condition may be an intriguing premise. We therefore would like to introduce the system of $NCH_3$ adsorbing on Cu(110) surface. In this paper, two distinct STM images of $NCH_3$ on adatom-involved Cu(110) surface have been determined using first-principles calculation. They are respective of zigzag and rectangle structures. The well-characterized zigzag structure which has been reported in our previous literatures is the preference in experimental observation over the rectangle structure. Our calculation results suggest that the existences of the two structures are attributed to various electronic configurations of $NCH_3$. $NCH_3$ adsorbed on Cu(110) surface is stable on its singlet state arranged as the zigzag structure; however, it is likely to be excited upon irradiation of light to produce triplet state. The occurrence of intersystem crossing coincident with corresponding geometrical variation which lead to the rectangle structure. The energy difference is 0.50 eV between two singlet and triplet $NCH_3$ molecules in a p(2x3) unit cell with 8 Cu layers and 4 adatoms. Moreover, triplet $NCH_3$ possessing reactive radicals perhaps triggers specific chemical reactions. On triplet state, $NCH_3$ can isomerize to $HNCH_2$. Next, the triplet $HNCH_2$ can bond with CO molecule to form $HNCH_2CO$, a middle fragment of glycine. The whole reaction is endothermic of 0.27 eV. The reaction is limited to occur at a high temperature region, for example near a terrestrial planetary in solar system. The system is manipulatble to extendedly study ISM reaction on dust surface.

**Figure captions**

Figure 1. Relative energy of isomerization of triplet and singlet NCH$_3$ to HNCH$_2$. * are from ref. [37] calculated by self-consistent field level of theory with double-zeta basis set. # is from ref. [38] calculated by CASPT2 level with cc-pVDZ basis set.

Figure 2. Top and lateral view of optimized geometries of NCH$_3$ adsorbed on Cu(110) surface in p(2x3) unit cell while (a) NCH$_3$ on singlet state and (b) NCH$_3$ on striplet state. In p(2x3) unit cell it consisted with two NCH$_3$ with four adatoms.

Figure 3. Diagrams of geometrical parameters of NCH$_3$ on Cu surface on (a) singlet state and (ref. [39]) (b) triplet state.

Figure 4. Experimental images and proposed structures of NCH$_3$ adsorbed on Cu(110). (a) Zig-zag STM pattern and NCH$_3$ on singlet state. (b) Rectangle STM pattern and NCH$_3$ on triplet state. We used VASP to check that the bright features are contributions from CH$_3$ groups [28]. Imaging conditions: sample V = 0.03 V, I = 0.13 nA. Image size: 4.5 nm × 4.5 nm.

Figure 5. (a) Diagram of chemical reactions of NCH$_3$ with CO to HNCH$_2$CO on Cu surface. Step I → II is isomerization of NCH$_3$ to HNCH$_2$. Step II → III is H=NCH bonding with CO to form HNCH$_2$CO, the middle fragment of glycine. (b) Potential energy curve of the chemical reactions on singlet state. (c) Potential energy curve of the chemical reactions on triplet state.

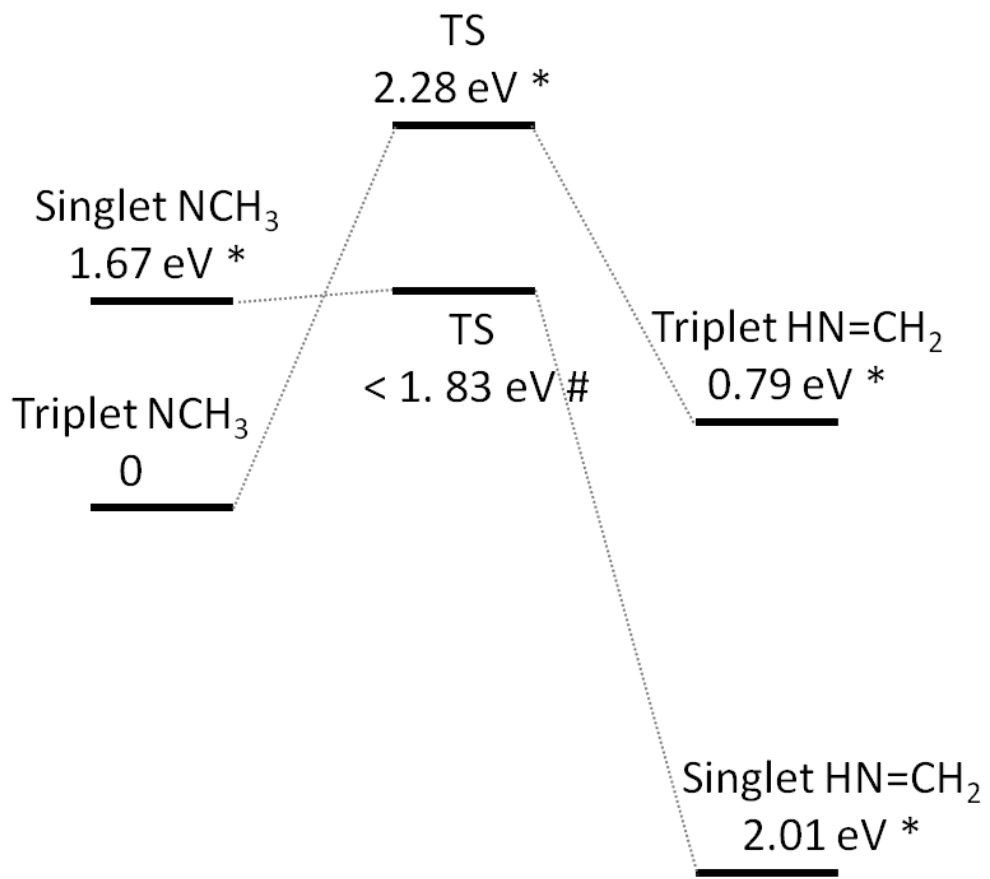

Figure 1

(a) Singlet state 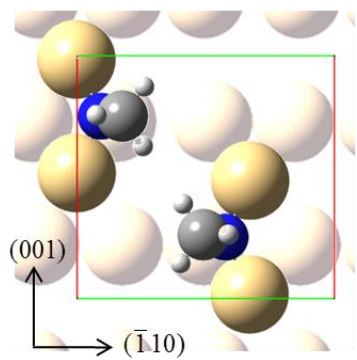

(b) Triplet state 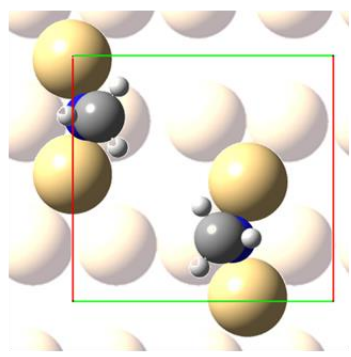

E = -0.50 eV

E = 0 eV

Figure 2

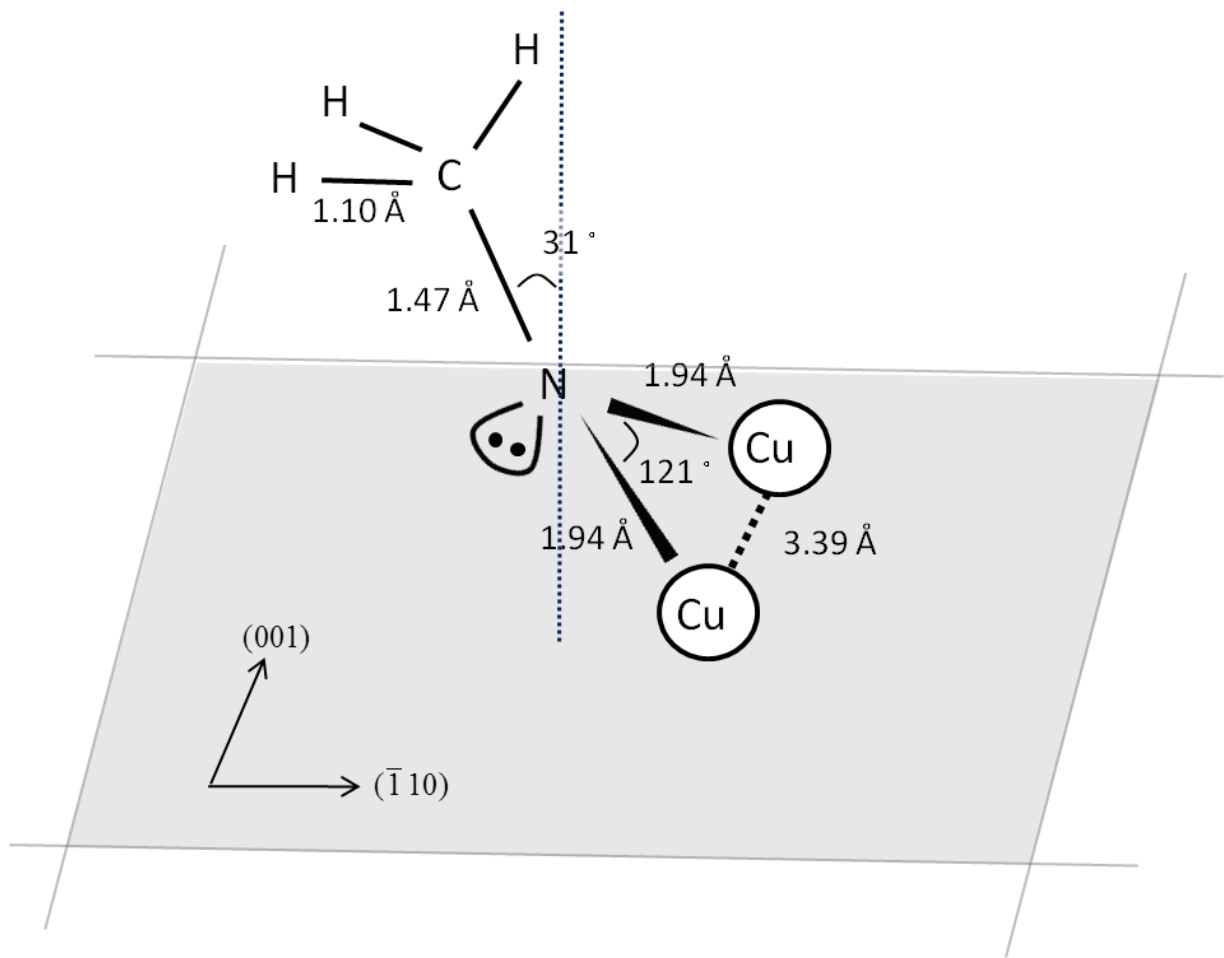

Figure 3(a)

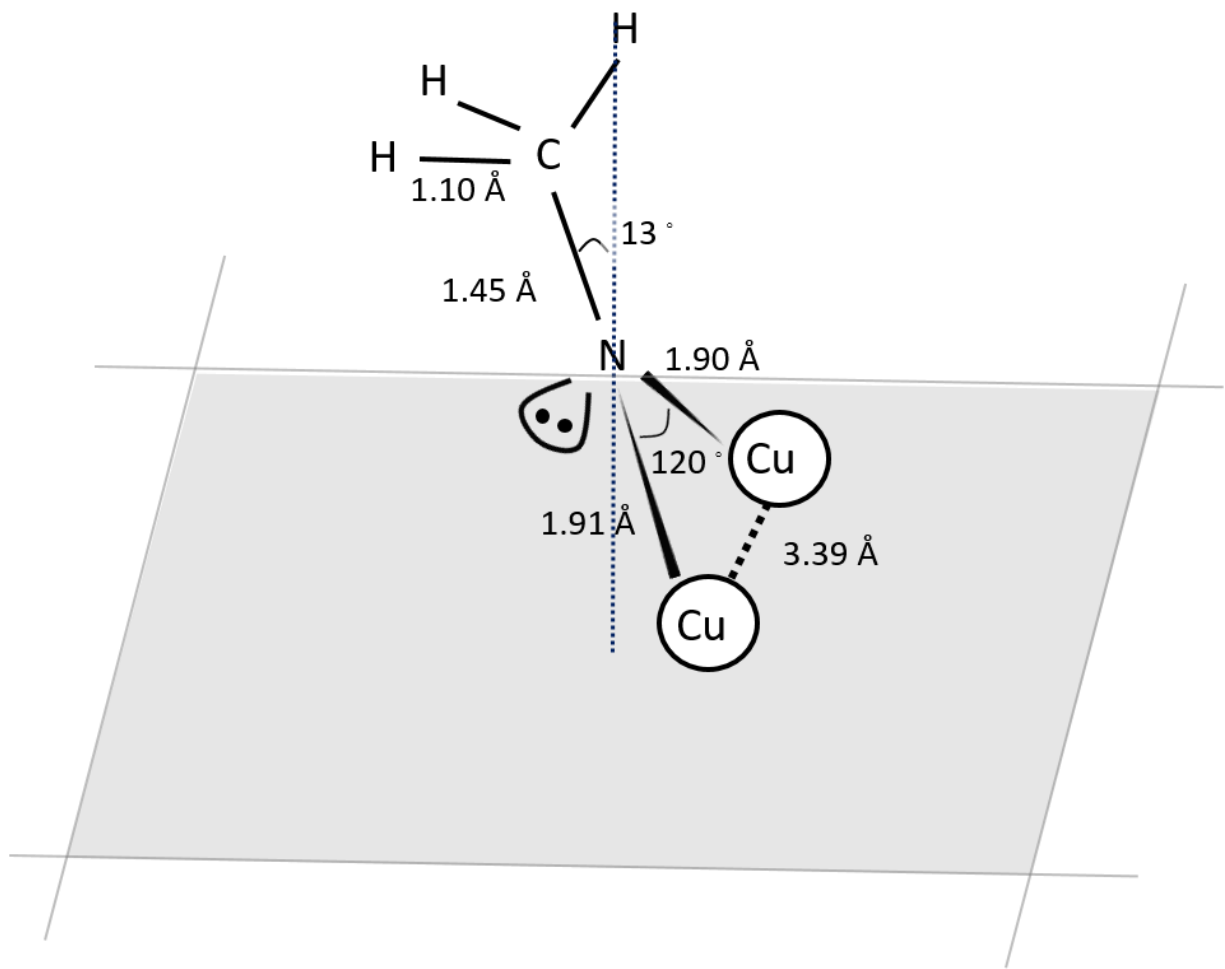

Figure 3(b)

(a) Singlet state 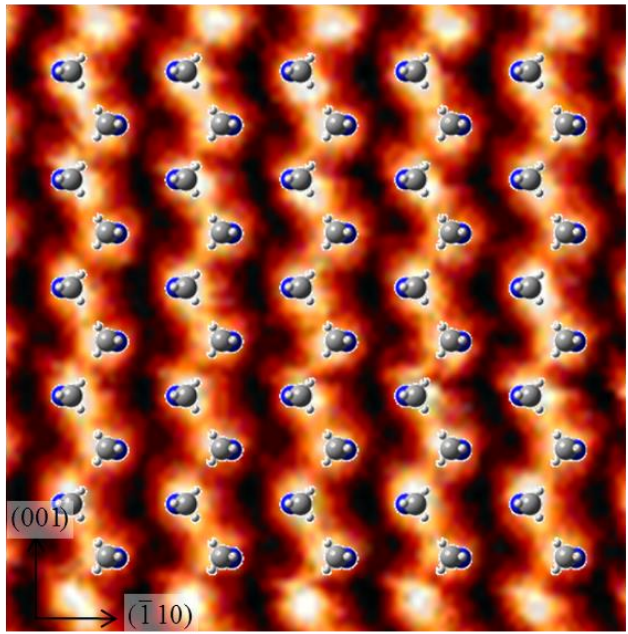   (b) Triplet state 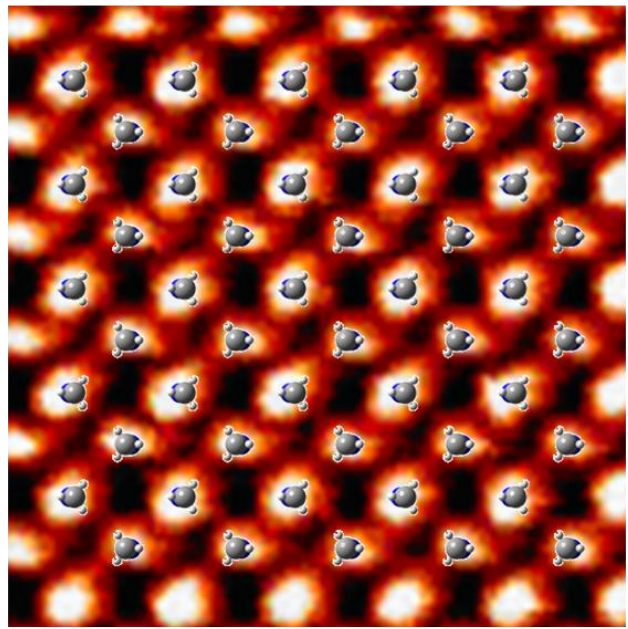

Figure 4

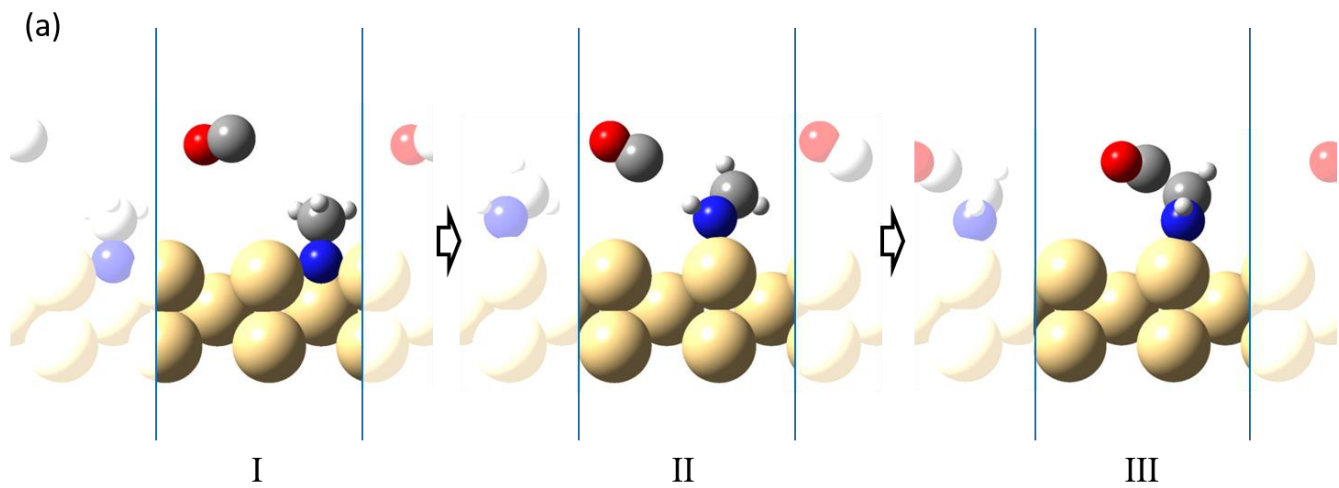

        I                      II                      III

Figure 5(a).

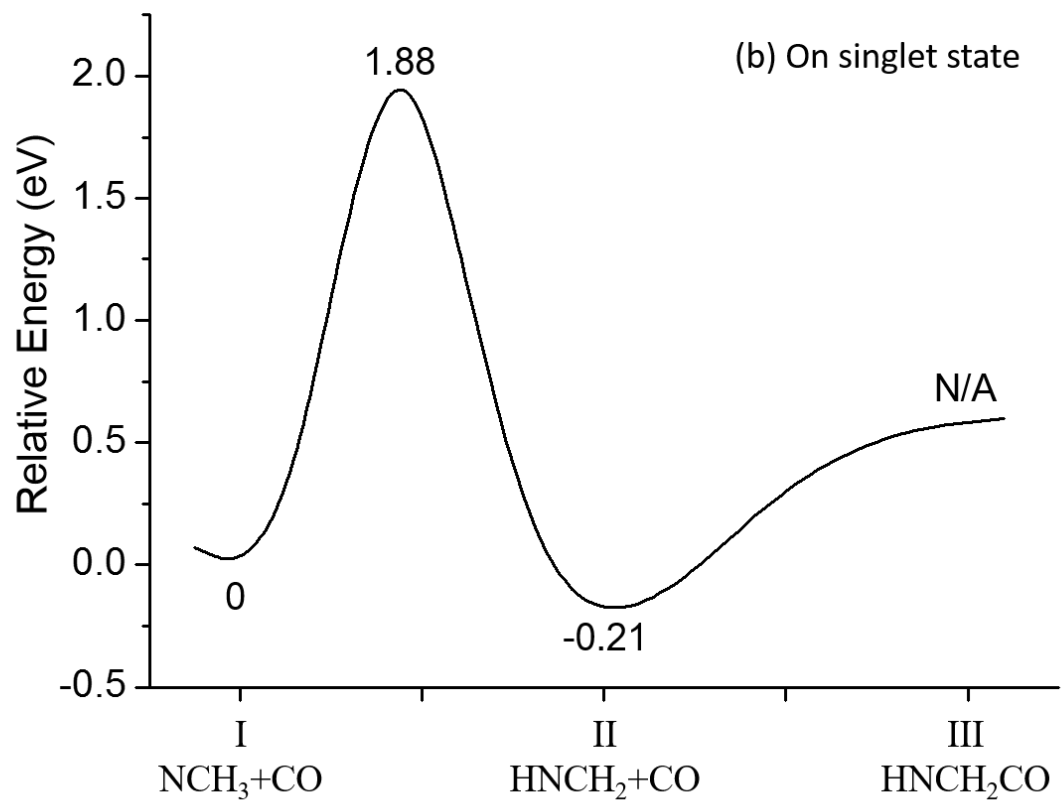

Figure 5(b).

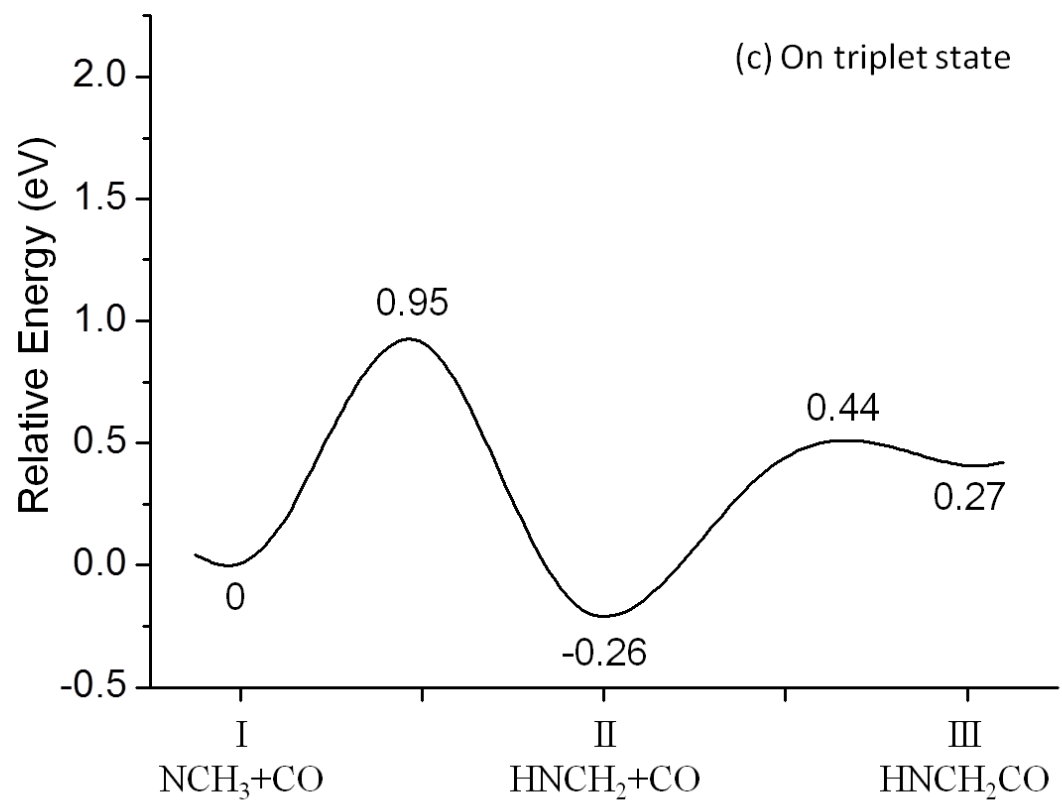

Figure 5(c).

# REFERENCES

[1] Eirc Herbst, The Chemistry of Interstellar Space. *Chem. Soc. Rev.* 2001, 30, 168–176.

[2] Eric Herbst, Chemistry in the Interstellar Medium. *Annu. Rev. Phys. Chem.* 1995, 46, 27–53.

[3] Jason A. Cardelli, Blair D. Savage, Dennis C. Ebbets, Interstellar Gas Phase Abundance of Carbon, Oxygen, Nitrogen, Copper, Gallium, Germanium, and Krypton toward Zeta Ophiuchi. *The Astrophysical Journal Letters* 1991, 383, L23–L27.

[4] Donatella Romano, Fransesca Matteucci, Contrasting Copper Evolution in ω Centauri and the Milky Way. *Monthly Notices of the Royal Astronomical Society Letter*. 2007, 378 (1), L59–L63

[5] Sheo S. Prasad, Shankar P. Tarafdar, UV Radiation Field inside Dense Clouds: Its Possible Existence and Chemical Implications. *The Astrophysical Journal Letters* 1983, 267, 603–609.

[6] R. Gredel, S. Lepp, A. Dalgarno, Eric Herbst, Cosmic-Ray-Induced Photodissociation and Photoionization Rates of Interstellar Molecules. *The Astrophysical Journal Letters* 1989, 347, 289–293.

[7] Pascale Ehrenfreund, Steven B. Charnley, Organic Molecules in the Interstellar Medium, Comets, and Meteorites: A Voyage from Dark Clouds to the Early Earth. *Annual Review of Astronomy and Astrophysics* 2000, 38, 427–483.

[8] J. R. Cronin, S. Chang, In The Chemistry of Life's Origins. 1993, 209–258. (Eds. J. M. Greenberg, C. X. Mendoza-Gomez, V. Pirronello.) Kluwer Academic Publishing, Dordrecht, The Netherlands.

[9] William M. Irvine, Extraterrestrial Organic Matter. A review. *Origin Life Evoution of the Biosphere.* 1998, 28, 365–383.

[10] A. Belloche, K. M. Menten, C. Comito, H. S. P. Müller, P. Schilke, J. Ott, S. Thorwirth, and C. Hieret, Detection of Amino Acetonitrile in Sgr B2(N). *Astronomy & Astrophysics* 2008, 482, 179–196.

[11] J. Oro, Comets and the Formation of Biochemical Compounds on the Primitive Earth. *Nature* 1961, 190, 389–390.

[12] C. F. Chyba, P. J. Thomas, L. Brookshaw, C. Sagan, Cometary Delivery of Organic Molecules to the Early Earth. *Science* 1990, 249, 366–373.

[13] A. H. Delsemme, Cometary Origin of the Biosphere: A Progress Report *Adv. Space Res.* 1995, 15, 49–57.

[14] P. Ehrenfreund, D. P. Glavin, O. Botta, G. Cooper, J. L. Bada, Extraterrestrial Amino Acids in Orgueil and Ivuna: Tracing the Parent Body of CI Type Carbonaceous Chondrites. *Proc. Natl. Acad. Sci.* 2001, 98, 2138–2141.

[15] Yi-Jehng Kuan, Steven B. Charnley, Hui-Chun Huang, Wei-Ling Tseng, Zbigniew Kisiel, Interstellar Glycine. *The Astrophysical Journal* 2003, 593, 848–867.

[16] L. E. Snyder, F. J. Lovas, J. M. Hollis, D. N. Friedel, P. R. Jewell, A. Remijan, V. V. Ilyushin, E. A. Alekseev, and S. F. Dyubko, A Rigorous Attempt to Verify Interstellar Glycine. *The Astrophysical Journal* 2005, 619 (2), 914–930.

[17] Keith A. Kvenvolden, James Lawless, Katherine Pering, Etta Peterson, Jose Flores, Cyril Ponnamperuma, Isaac R. Kaplan, Carleton Moore, Evidence for Extraterrestrial Amino-Acids and Hydrocarbons in the Murchison Meteorite. *Nature* 1970, 228 (5275), 923–926.

[18] M. P. Bernstein, J. P. Dworkin, S. A. Sandford, G.W. Cooper, L. J. Allamandola, Racemic Amino Acids from the Ultraviolet Photolysis of Interstellar Ice Analogues. *Nature* 2002, 416, 401–403.

[19] P. D. Holtom, C. J. Bennett, Y. Osamura, N. J. Mason, R. I. Kaiser, A Combined Experimental and Theoretical Study on the Formation of the Amino Acid Glycine ($NH_2CH_2COOH$) and its Isomer ($CH_3NHCOOH$) in Extraterrestrial Ices. *The Astrophysical Journal* 2005, 626, 940–952.

[20] A. Lafosse, M. Bertin, A. Domaracka, D. Pliszka, E. Illenberger, R. Azria, Reactivity Induced at 25 K by Low-Energy Electron Irradiation of Condensed $NH_3$–$CH_3COOD$ (1:1) Mixture. *Phys. Chem. Chem. Phys.* 2006, 8, 5564–5568.

[21] Chang-Woo Lee, Joon-Ki Kim, Eui-Seong Moon, Y. C. Minh, and Heon Kang, Formation of Glycine on Ultraviolet-Irradiated Interstellar Ice-Analog Films and Implications for Interstellar Amino Acids. *The Astrophysical Journal* 2009, 697, 428–435.

[22] Zanele P. Nhlabatsi, Priya Bhasia, Sanyasi Sitha, Possible interstellar formation of glycine from the reaction of $CH_2$=NH, CO and $H_2O$: catalysis by extra water molecules through the hydrogen relay transport *Phys. Chem. Chem. Phys.* 2016, 18, 375–381.

[23] Albert Rimola, Mariona Sodupe, and Piero Ugliengo, Computational Study of Interstellar Glycine Formation Occurring at Radical Surfaces of Water-Ice Dust Particles. *The Astrophysical Journal* 2012, 754, 24.

[24] Gaston Berthier, Chemical Reactivity in Interstellar Space, in Introduction to Advanced Topics of Computational Chemistry 2003, 335–354. L. A. Montero, L. A. Díaz and R. Bader (eds.)

[25] Zhai, R. S.; Chan, Y. L.; Hsu, C. K.; Chuang, P.; Klauser, R.; Chuang, T. J. Generation and Spectroscopic Characterization of Methylnitrene Diradicals Adsorbed on the Cu(110) Surface. *ChemPhysChem.* **2004**, *5*, 1038–1041.

[26] Pai, W. W.; Chan, Y. L.; Chuang, T. J. Chemisorption of Methyl ($CH_3$) and Methylnitrene ($NCH_3$) Radicals on Cu Surfaces Studied by STM and LEED. *Chin. J. Phys.* **2005**, *43*, 212–218.

[27] Pai, W. W.; Chan, Y. L.; Chang, S. W.; Chuang, T. J.; Lin, C. H. Adsorption of Methyl ($CH_3$) and Methylnitrene ($NCH_3$) Radicals on Cu Surfaces Studied by Scanning Tunneling Microscopy. *Jpn. J. Appl. Phys.* **2006**, *45*, 2372–2376.

[28] Chen, P. T.; Pai, W. W.; Chang, S. W.; Hayashi, M. Scanning Tunneling Microscopy and Density Functional Theory Studies of Adatom-Involved Adsorption of Methylnitrene on Copper (110) Surface. *J. Phys. Chem. C* **2013**, *117*, 12111–12116.

[29] Kresse, G.; Hafner, J. Ab Initio Molecular Dynamics for Open-Shell Transition Metals. *Phys. Rev. B* **1993**, *48*, 13115–13118.

[30] Kresse, G.; Furthmüller, J. Efficient Iterative Schemes for Ab Initio Total-Energy Calculations Using a Plane-Wave Basis Set. *Phys. Rev. B* **1996**, *54*, 11169–11186.

[31] Kresse, G.; Furthmüller, J. Efficiency of Ab Initio Total Energy Calculations for Metals and Semiconductors Using a Plane-Wave Basis Set. *Comput. Mat. Sci.* **1996**, *6*, 15–50.

[32] Perdew, J. P.; Wang, Y. Accurate and Simple Analytic Representation of the Electron-Gas Correlation Energy. *Phys. Rev. B* **1992**, *45*, 13244–13249.

[33] Monkhorst, H. J.; Pack, J. D. Special Points for Brillouin-Zone Integrations. *Phys. Rev. B* **1976**, *13*, 5188–5192.

[34] G. Henkelman and H. Jónsson, A Climbing Image Nudged Elastic Band Method for Finding Saddle Points and Minimum Energy Paths. *J. Chem. Phys.* 2000, 113, 9901–9904.

[35] G. Henkelman and H. Jónsson, Improved Tangent Estimate in the Nudged Elastic Band Method for Finding Minimum Energy Paths and Saddle Points. *J. Chem. Phys.* 2000, 113, 9978–9985.

[36] Yarkony, D. R.; Schaefer III, H. F.; Rothenberg, S. X $^3A_2$, a $^1E$, and b $^1A_1$ Electronic States of Methylnitrene. *J. Am. Chem. Soc.* **1974**, *96*, 5974–5977.

[37] Demuynck, J.; Fox, D. J.; Yamaguchi, Y.; Schaefer III, H. F. Triplet Methylnitrene: An Indefinitely Stable Species in the Absence of Collisions. *J. Am. Chem. Soc.* **1980**, *102*, 6204–6207.

[38] Kemnitz, C. R.; Ellison, G. B.; Karney, W. L.; Borden, W. T. CASSCF and CASPT2 Ab Initio Electronic Structure Calculations Find Singlet Methylnitrene Is an Energy Minimum. *J. Am. Chem. Soc.* **2000**, *122*, 1098–1101.

[39] Po-Tuan Chen, Woei Wu Pai, Michitoshi Hayashi, A Minimal Cluster Model of Valence Electrons in Adatom-Assisted Adsorbed Molecules: NCH$_3$/Cu(110) and OCH$_3$/Cu(110). *J. Phys. Chem. C* 2014, 118, 9443−9449